\documentclass[aps,twocolumn,groupedaddress,showpacs]{revtex4}
\usepackage{graphics}
\begin{document}
\newcommand{\beq}{\begin{equation}}
\newcommand{\eeq}{\end{equation}}
\bibliographystyle{apsrev}

\title{Dynamics of Rolling Massive Scalar Field Cosmology}

\author{Pravabati Chingangbam}
\email{prava@mri.ernet.in}
\affiliation{Harish-Chandra Research Institute, Chhatnag Road, Jhunsi,
Allahabad-211019, India.}
\author{Tabish Qureshi}
\email{tabish@jamia-physics.net}
\affiliation{Department of Physics, Jamia Millia Islamia, New Delhi-110025,
India.}

\begin{abstract}
We study the inflationary consequences of the rolling massive scalar field 
in the braneworld scenario with a warped metric. We find that in order to 
fit observational constraints the warp factor must be tuned to be $< 
10^{-3}$. We also demonstrate the inflationary attractor behavior of the 
massive scalar field dynamics both in the standard FRW case as well as in 
braneworld scenario. 

\end{abstract}
\maketitle

\section{Introduction}

The inflationary paradigm \cite{inflation}, despite overwhelming support 
from observational 
data \cite{wmap}, is still ad hoc. There still is no conclusive model 
supported by a 
fundamental theory which provides a mechanism for its realization. It is 
hoped that M/string theory may provide the theoretical framework for 
realization of inflation and there has recently been a lot of activity 
towards construction of models of inflation in string theory. See 
\cite{quevedo} for a review. Originating from the work of Sen \cite{sen}, 
the possibility of the tachyon field being a candidate for the 
inflaton has 
been extensively studied \cite{cosmo}. The tachyon action is of the Dirac 
Born Infeld form \cite{dbi} which leads to an equation of state 
interpolating between  
-1 at early times and 0 at late times. This  
suggests the possibility that the tachyon can play the role of  
the inflaton in early times and  dark matter at late times. 
However tachyonic inflation is found to be plagued by serious 
difficulties, namely, large density perturbations, problem with reheating 
and formation of caustics \cite{linde}. 

In a recent paper \cite{sami} Garousi, Sami and Tsujikawa put forward 
an inflationary model 
provided by a DBI type effective field theory of rolling massive 
scalar field on a $D$ brane or anti-D brane, obtained from string theory 
\cite{garousi}. Not surprisingly, as in the case of tachyon inflation 
the amplitude of density perturbations obtained are too large. However by 
considering the $D$ brane in a warped background, the warping factor is 
introduced in the model as a parameter that can be tuned to get 
acceptable density perturbations. The authors demonstrate that the 
inflationary observables in the model, such as the spectral index and 
tensor-to-scalar ratio, can be fitted within the 
observational constraints, thus making it a viable model for inflation. 
Typically a very small value of the warp factor, of the order of 
$10^{-8}$ is 
required to fit the cosmological constraints. Fluxes are the sources of 
warping in flux compactification and it has been shown that a 
very small value of $\beta $ is 
indeed possible for suitable choices of fluxes \cite{warp}.

In this paper we study the cosmological consequences of the rolling 
massive scalar field on a (4+1)-dimensional brane world scenario. The 
motivation is to see whether the enhancement in inflation 
due to the brane correction term in the Friedmann equation renders it 
possible to have a viable inflationary model without tuning the warp 
factor 
to be small.  We find that a small value of the warp factor is still a 
necessity so as to get correct amount of density 
perturbations. However its order of magnitude must be $10^{-3}$ in 
contrast to the $10^{-8}$ in standard FRW cosmology. 

We also study the phase space behavior of the rolling 
massive scalar field in standard FRW cosmology and in the braneworld scenario.
We demonstrate that the phase trajectories rapidly converge towards 
the fixed point thereby exhibiting attractor behavior.

\section{Review of massive rolling scalar field inflation in warped 
compactification}

Consider a BPS-D3-brane in type IIB string theory where the six extra 
dimensions are compactified. 
The effective action for the rolling massive scalar field obtained from 
string theory is given by
\beq
S = -\int d^4x V(\phi)\sqrt{-\rm{det}\big(\eta_{ab}
+\partial_a\phi\partial_b\phi\big)}
\eeq
where $\phi$ has the unit of $\sqrt{\alpha'}$, $\ \alpha'$ being the 
string length.
The mass of the field $\phi$ is given by $m^2=(n-1)/\alpha'$, 
$n$ being an integer and so for $n\ge 2$ the
scalar fields are massive. 

The potential of the scalar field, to fourth order in $\phi$, was derived 
\cite{garousi} to be
\beq
V(\phi) = \tau_3\Big(1+ {1\over 2}m^2\phi^2 + {1\over
8}m^4\phi^4+\ldots\Big)
\eeq
which in ref \cite{sami} is speculated to have the closed form
\beq
V(\phi) = \tau_3 e^{{1\over 2}m^2\phi^2}
\eeq

Consider the warped metric 
\beq
ds_{10}^2 = \beta(y_i)\eta_{ab}dx^a dx^b + \beta^{-1}(y_i) g_{ij}dy^idy^j
\eeq
where $\beta(y_i)$ is the warping factor depending only on the coordinates 
of the compact directions $y_i$ and $\tilde{g}_{ij}$ is the metric on the 
compact space. Let us consider a scenario in which the brane can move in 
the compact space thereby reducing its tension. This corresponds 
physically to the situation where the warp factor $\beta$ is small at 
some point on the compact space. The action for a massive scalar field 
rolling on the $D_3$-brane minimally coupled to gravity is then given by 
the DBI form  
\beq
S = -\int d^4x
\beta^2V(\phi)\sqrt{-\rm{det}\big(\eta_{ab}
+\beta^{-1}\partial_a\phi\partial_b\phi\big)}
\eeq
Redefining the scalar field as $\phi\rightarrow \sqrt{\beta}\phi$
we get 
\beq
S = -\int d^4x
V(\phi){\sqrt{-\rm{det}\big(\eta_{ab}+\partial_a\phi\partial_b\phi
\big)}}
\eeq
with the redefined potential
\beq
V(\phi) = V_0e^{{1\over 2}m^2\beta\phi^2},\quad 
V_0=\beta^2\tau_3\label{potential}
\eeq
In a spatially flat FRW background with a scale factor $a$, the energy 
density $\rho$ and the pressure $p$ of the field are given by
\beq
\rho = {{V}\over{\sqrt{1-\dot{\phi}^2}}} 
\eeq
\beq
p = -V\sqrt{1-\dot{\phi}^2}
\eeq

The inflaton equation of motion which follows from (\ref{potential}) is 
\beq
{{\ddot{\phi}}\over{1-\dot{\phi}^2}} + 3H\dot{\phi} \label{motion}
+{{V'}\over V} = 0 
\eeq
where the prime denotes differentiation with respect to $\phi$. 
The Friedmann equation is 
\begin{equation} 
H^2 = {{\kappa^2}\over
3}{V\over{\sqrt{1-\dot{\phi}^2}}} 
\end{equation} 
The slowroll parameters are 
\begin{equation}
\epsilon = {{M_P^2}\over 2}{{V'^2}\over{V^3}}, \qquad
\eta = -2\epsilon + M_P^2 {{V''}\over{V^2}}
\end{equation}

For the potential given by (\ref{potential}) they take the form
\beq
\epsilon = {{m^4M_P^2}\over{2\tau_3}}\phi^2 e^{-{1\over 2}m^2\beta\phi^2}
\eeq
\beq
\eta = {m^2 M_P^2 \over \beta\tau_3} e^{-{1\over 2}m^2\beta\phi^2} 
\eeq
The slow roll condition $\epsilon<1$ and $|\eta|<1$ is trivially satisfied 
by choosing sufficiently large initial $\phi$.
The number of e-folding during inflation is given by
\beq
N = \int_{t_i}^{t_{\rm{end}}} H dt.
\eeq
By choosing initial $\phi$ large enough it is always possible to get 
sufficient number of efoldings.

The amplitude of the density perturbation is given by \cite{hwang}
\beq
\mathcal{P}_s \sim  {{H^4}\over{(2\pi\dot{\phi})^2 V}} 
\end{equation}
which in the slow roll approximation becomes
\beq
\mathcal{P}_s \simeq {1\over{12\pi^2M_P^2}}\Big({{V^2}\over{V'}}\Big)^2
\eeq

and for the potential (\ref{potential}) it becomes 
\beq
\mathcal{P}_s \simeq {{\beta^2}\over{48\pi^2\epsilon^2}}
         {{m^6\phi^2}\over{M_P^2}}
\eeq

Observational constraints demand that $\mathcal{P}_s \simeq 10^{-9}$, from 
which we can get an estimate on the range of $\beta$. Taking 
$m^2\beta\phi^2\sim 1$ and $\epsilon\sim 1$ and 
putting $ m/M_P\sim 1$ we get
$\beta< 10^{-8}$. Of course for getting the correct estimate of 
$\beta$ we must put the values of $\phi$ and $\epsilon$ at fifty 
efolds or so, in which case $\beta$ will have to be much less than the 
above 
estimate.  Here we are only highlighting the fact that $\beta$ has to be 
tuned to be much less than one to fit the observational constraints. 
Thus we see that in the absence of warping, that is for $\beta = 1$, 
it is impossible to 
get a viable model of inflation from the rolling massive scalar field.

\section{Rolling scalar field on the brane}

In the braneworld scenario the prospects of inflation become highly 
favourable due to high energy corrections to the Friedmann equation.
\beq
H^2 = {1\over{3M_P^2}}\rho \Big(1+{{\rho}V\over{\lambda_b}}\Big)
\eeq
where $\lambda_b$ is the brane tension.
In the high energy limit we have $\rho/\lambda_b\gg 1$ and inflation 
occurs as 
long as $\dot{\phi}^2<1/3$. In this section we analyze inflation for the 
rolling massive scalar field in the braneworld scenario to see whether we still 
require to tune the warp factor to get the right amount of density 
perturbations. 

The slow roll parameters take the form
\beq
\epsilon = 12M_P^2\lambda_b \big({{V'}\over{V^2}}\big)^2
\eeq
\beq
\eta = -6M_P^2\lambda_b {{V''}\over{V^3}}
\eeq
The slow roll condition can be satisfied by choosing large initial $\phi$. 
The number of efolds is
\begin{eqnarray*}
N &=& -{1\over{2M_P^2\lambda_b}}\int_{\phi_i}^{\phi_f}
         {{V^3}\over{V'}}d\phi\\
{} &=& -{1\over 4} {{V_0^2}\over{M_P^2\lambda_b m^2\beta}}
            \int_{x_i}^{x_f}{{e^x}\over{x}}dx
\end{eqnarray*}
where $x=m^2\beta\phi^2$. Again sufficient number of efoldings can always 
be obtained by choosing large initial $\phi$. 
The amplitude of density perturbations is
\begin{eqnarray*}
\mathcal{P}_s &=& {9\over{4\pi^2(6M_P^2\lambda_b)^3}}
                  {{V^7}\over{V'^2}} \\
{}            &=& {3\over{2\pi^2}}{{\beta^2V_0}\over{M_P^2\lambda_b}}
                    {{m^4\phi^2}\over{\epsilon^2}}
                         e^{{1\over 2}m^2\beta\phi^2}
\end{eqnarray*}
To estimate $\beta$ we put $m^2\beta\phi^2\sim 1$ and $\epsilon \sim 
1$ and 
$\tau_3/(m^2M_P^2) \sim 1$ which gives
\beq
\mathcal{P}_s \simeq {3\over{2\pi^2}}{{\beta^3}\over{\lambda_b}}m^4
\eeq
Putting $\mathcal{P}_s\simeq 10^{-9}$ 
we get 
\beq
\beta^3 \le {{\lambda_b}\over{m^4}} 10^{-9}
\eeq
Since the brane tension $\lambda_b$ must be $\ll M_P^4$ we have
\beq
\beta^3 \ll \big({{M_P}\over{m}}\big)^4 10^{-9}
\eeq
which on taking $m/M_P\sim 1$ gives $\beta\ll 10^{-3}$. Thus in the 
braneworld scenario we still crucially require small value for 
$\beta$ despite the enhanced value of the Hubble parameter $H$ coming 
from the brane correction.

\section{Inflationary attractor for DBI massive scalar field}

In this section we demonstrate the attractor behavior of rolling scalar field 
inflation cosmology in standard FRW and brane world 
scenarios. 
Introducing the dimensionless variables
\beq
x\equiv\dot{\phi},\quad y\equiv {V\over{V_0}},\quad \tau\equiv mt
\eeq
we can write Eq.(\ref{motion}), subject to the constraint imposed 
by the Friedmann equation, as the following two coupled equations
\begin{eqnarray}
x' &=& -\sqrt{{3V_0}\over{m^2M_P^2}} \sqrt{y} x(1-x^2)^{3/4} - 
\sqrt{2\beta\ln y}(1-x^2)\nonumber\\
y' &=& \sqrt{2\beta} xy\sqrt{\ln y}\label{xmotion}
\end{eqnarray}
where prime denotes differentiation with respect to $\tau$ in this section. 
We get the fixed points $(0,1)$ and $(\pm 1,1)$. The latter two points are 
not physically interesting because they can be 
removed by another choice of 
coordinates, say $x\equiv \dot\phi$ and $y\equiv \phi$.  In order 
to see the stability of the critical point (0,1) we perturb about the 
fixed point 
\begin{eqnarray*}
x &=& \epsilon\\
y &=& 1+ \delta
\end{eqnarray*}
where $\epsilon$ and $\delta$ are infinitesimally small. Putting them in 
eqs. (\ref{xmotion}) and keeping only lowest order terms we get
\begin{eqnarray*}
\epsilon' &=& -\sqrt{{{3V_0}\over{m^2M_P^2}}} \epsilon 
             +\sqrt{2\beta\delta}\\
\delta' &=& \sqrt{2\beta\delta} \epsilon
\end{eqnarray*}
Defining $\gamma^2/2 \equiv \delta$ we can linearize the above 
equations to get  
\begin{eqnarray*}
\epsilon' &=& -\sqrt{{{3V_0}\over{m^2M_P^2}}} \epsilon
             +\sqrt{\beta\gamma}\\
\gamma' &=& \sqrt{\beta}\epsilon
\end{eqnarray*}
which can be solved in matrix form to get
\beq
\left(\begin{array}{c}
\epsilon  \\
\gamma
\end{array} \right)
= 
\left(\begin{array}{c}
\epsilon_0  \\
\gamma_0
\end{array} \right)
e^{R\tau}
\eeq

where $R$ is given by
\beq
\left(\begin{array}{cc}
-\sqrt{{{3V_0}\over{m^2M_P^2}}} & -\sqrt{\beta} \\
1 & 0
\end{array} \right)
\eeq
having the eigenvalues 
\beq
\lambda_{\pm} = -{1\over 2} \sqrt{{3V_0}\over{m^2M_P^2}} 
                  \pm \sqrt{ {{3V_0}\over{m^2M_P^2}} -4\sqrt{\beta} }
\eeq

It is clear that the real part of $\lambda_{\pm}$ is always negative and 
so the fixed point (0,1) is stable. As can be seen from the numerical 
solution of Eq.(\ref{xmotion}) 
shown in Fig.(1) trajectories originating at any initial point on 
the phase space end up at the stable fixed point (0,1). 

We next study the phase space behavior in the braneworld scenario. 
On the brane Eqs.(\ref{xmotion}) gets modified to
\begin{eqnarray}
x' &=& -\sqrt {{3V_0}\over{m^2M_P^2}} \sqrt{{y}\over{\sqrt{1-x^2}}}
      \Big( 1 + {{V_0}\over{\lambda_b}} {{y}\over{\sqrt{1-x^2}}} \Big) 
           x(1-x^2)\nonumber\\
{}  &{}& - \sqrt{2\beta\ln y}(1-x^2)
\end{eqnarray}

The fixed points are the same as the previous case. The stability 
analysis can be carried out as before. The phase 
trajectories are shown in fig. (2). All trajectories are seen to move  
toward the stable fixed point (0,1). 

From a comparion of the trajectories in  Figs. (1) and (2) we see that 
they approach the 
stable fixed point (0,1) faster in the braneworld case. This is a 
consequence of the enhancement of the friction term in the equation of 
motion of $\phi$ due to the high energy correction in the Friedmann 
equation. 

\begin{figure}
\resizebox{!}{2.5in}{\includegraphics{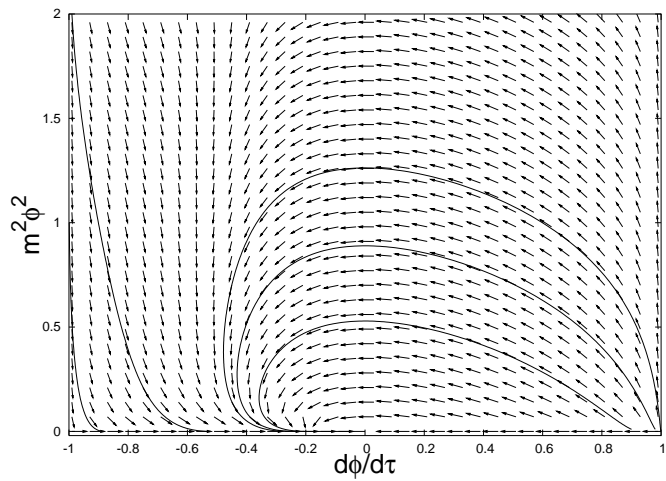}}
\caption{ Phase plot of rolling massive scalar field in standard FRW 
cosmology. Trajectories approach the stable fixed point (0,1).}
\end{figure}

\begin{figure}
\resizebox{!}{2.5in}{\includegraphics{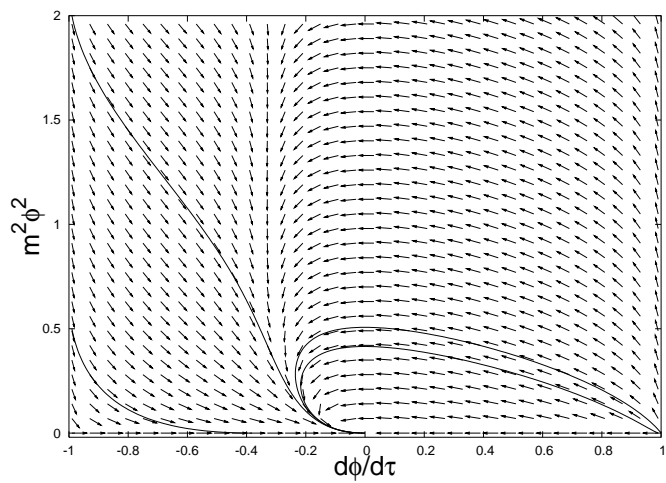}}
\caption{ Plot of phase trajectories for rolling massive scalar field in 
braneworld. As in the previous figure, trajectories approach the stable 
fixed point (0,1) but at a faster rate.}
\end{figure}



\section{Conclusion}

In this paper we have shown that for the rolling massive scalar field to
be a good candidate for the inflaton in a brane world scenario, tuning of
the warping factor to be small is still crucially necessary, despite the
enhancement of inflation coming from the brane correction term. 
The order of
magnitude of the warping factor is however much improved. The difficulty
with tachyonic inflation, namely that of large density perturbations, in
conventional compactification is also faced by the rolling massive scalar
field inflation even though the potentials in the two models are vastly
different.  The reason is, the two parameters in the model, namely, the
string coupling $g_s$ and the compactification volume $v$ are constrained
by string theory to be $g_s<1$ and $v\gg 1$, which is incompatible with
the requirement for small density perturbations. It must be pointed out
that while warped compactification solves the problem for the rolling
massive scalar field it does not do so for the tachyon. The slow roll
condition for the tachyon, in conventional compactification, is given in
terms of $g_s$ and $v$ as roughly $g_s/v \gg 127$. In the warped
compactification this is modified to $g_s \beta /v \gg 127$ which cannot 
be satisfied for small $\beta$. However, in a warped compactification and 
varying $g_s$ the tachyon may still be a a good candidate 
for the inflaton as shown recently in \cite{joris} . 

We have further studied the dynamics of the rolling scalar field. In 
particular we have demonstrated the attractor behavior of inflation driven 
by this field in standard FRW cosmology and brane world cosmology. The 
trajectories show attractor behavior and rapidly converge towards the 
fixed point.   
Thus the evolution of the field, and consequently the onset of inflation, 
is independent of the initial conditions one may choose on the phase 
space, implying that the predictions of the physical observables, such as 
the amplitude of density perturbations are indeed independent of the 
initial conditions.

\begin{acknowledgments}
We thank M. Sami for valuable discussions. The authors are grateful to 
IUCAA, Pune for a visit, during which the present work was 
carried out.
\end{acknowledgments}


\begin{thebibliography}{99}
\bibitem{inflation} A. Guth, Phys. Rev. D {\bf 23}, 347 (1981);
A. D. Linde, {\em Particle Physics and Inflationary Cosmology},
           Harwood Chur (1990);
E. W. Kolb and M. S. Turner, {\em The Early Universe}, Addison--Wesley,
        Redwood City (1990);
A. R. Liddle and D. H. Lyth, {\em Cosmological Inflation
  and Large-Scale Structure}, Cambridge University Press (2000).

\bibitem{wmap}
C. L. Bennett et al., Astrophys. J. Suppl. {\bf{148}}, 1 (2003);
D. N. Spergel et al., Astrophys. J. Suppl. {\bf{148}}, 175 (2003);
H. V. Peiris et al., Astrophys. J. Suppl. {\bf{148}}, 213 (2003);
G. Hinshaw et al., Astrophys. J. Suppl. {\bf{148}}, 135 (2003).



\bibitem{quevedo} F. Quevedo, Class. Quant. Grav. {\bf{19}}, 5721 (2002)
[hep-th/0210292]

\bibitem{sen} A. Sen, JHEP {\bf{0204}}, 048 (2002), [hep-th/0203211]; 
A. Sen, JHEP 0207 (2002) 065, [hep-th/0203265]

\bibitem{cosmo} 
S. Alexander, hep-th/0105032;
A. Mazumdar, S. Panda and A. Perez-Lorenzana, Nucl. Phys. B {\bf 614},
101 (2001) [hep-ph/0107058];
G. Gibbons, Phys. Lett. B {\bf{537}}, 1 (2002), hep-th/0204008;
M. Fairbairn and M. Tytgat, Phys.Lett.B546:1-7, (2002), hep-th/0204070;
S. Mukohyama, Phys.Rev.D66:123512, (2002), hep-th/0208094;
A. feinstein, Phys.Rev.D66:063511, (2002), hep-th/0204140; 
T. Padmanabhan, Phys.Rev.D66:021301, (2002), hep-th/0204150; 
A. Frolov, L. Kofman and A. Starobinsky, Phys.Lett.B545:8-16, (2002),  
hep-th/0204187; 
D. Choudhary, D. Ghoshal, D. Jatkar and S. 
Panda, Phys.Lett.B544:231-238, (2002), hep-th/0204204;
X. Li, J. Hao and D. Liu, Chin.Phys.Lett.19:1584,(2002), hep-th/0204252; 
G. Shiu and I. Wasserman, Phys.Lett.B541:6-15, (2002), hep-th/0205003; 
T. Padmanabhan and T. R. Choudhury, 
Phys.Rev.D66:081301, (2002), hep-th/0205055;
H. B. Benaom, hep-th/0205140; 
M. Sami, P. Chingangbam and T. Qureshi, Phys. Rev. {\bf{D
                66}}, 043530 (2002) [hep-th/0205179];
G. Shiu, S. H. Tye and I. Wasserman, 
Phys.Rev.D67:083517, (2003), hep-th/0207119; 
Y. S. Piao, R. G. Cai, X. m. Zhang and Y. Z. Zhang, 
Phys.Rev.D66:121301, (2002),  hep-ph/0207143;
X. z. Li, D. j. Liu and J. g. Hao, hep-th/0207146;
J. M. Cline, H. Firouzhahi and P. Martineau, JHEP 0211:041, (2002), 
hep-th/0207156;
G. Felder, L. Kofman and A. Starobinsky, JHEP 
0209:026, (2002), hep-th/0208019; 
B. Wang, E. Abdalla and R. K. Su, 
Mod.Phys.Lett.A18:31-40, (2003), hep-th/0208023; 
M. C. Bento, O. Bertolami and A. A. Sen, 
Phys.Rev.D67:063511, (2003), hep-th/0208124; 
Jian-gang Hao and Xin-zhou Li, Phys.Rev.D66:087301, (2002), 
hep-th/0209041; 
G. Shiu, hep-th/0210313;
Y. S. Piao, Q. C. Huang,  X. m. Zhang and Y. Z. Zhang, Phys. Lett. 
B{\bf{570}} 1 (2003), hep-th/0212219;
M. Sami, P. Chingangbam and T. Qureshi, Pramana 62:765, 
(2004), hep-th/0301140;
X. z. Li and X. h. Zhai, Phys. Rev. D{\bf{67}} 067501 (2003), 
hep-th/0301063; 
G. Gibbons, Class.Quant.Grav.20:S321-S346, (2003), hep-th/0301117; 
Z. K. Guo, Y. S. Piao, R. G. Cai and Y. Z. Zhang,  Phys. Rev. D{\bf{68}} 
043508 (2003), hep-th/0304236;
Y. Demasure and R. A. Janik, Phys.Lett.B578:195-202, (2004), 
hep-th/0305191; 
Z-K. Guo and H-S Zhang and Y-Z Zhang, Phys.Rev. D69 (2004) 063502 
hep-ph/0309163; 
M. Majumdar and A. C. Davis,
Phys.\ Rev.\ D {\bf 69}, 103504 (2004)
[arXiv:hep-th/0304226];
D. Choudhury, D. Ghoshal, D. P. Jatkar and S. Panda,
JCAP {\bf 0307}, 009 (2003)
[arXiv:hep-th/0305104];
Y. S. Piao and Y. Z. Zhang,
arXiv:hep-th/0307074;
L. R. W. Abramo and F. Finelli,
potential,''
Phys.\ Lett.\ B {\bf 575}, 165 (2003)
[arXiv:astro-ph/0307208];
M. C. Bento, N. M. C. Santos and A. A. Sen,
arXiv:astro-ph/0307292;
D.A. Steer and F. Vernizzi, hep-th/0310139;
V.Gorini, A.Y.Kamenshchik, U.Moschella and V.Pasquier,
arXiv:hep-th/0311111;
B.C.Paul and M.Sami,
arXiv:hep-th/0312081;
Ashoke Sen,
arXiv:hep-th/0312153;
J.M.Aguirregabiria and R.Lazkoz,
Mod.\ Phys.\ Lett.\ A {\bf 19}, 927 (2004)
[arXiv:gr-qc/0402060];
J.M.Aguirregabiria and R.Lazkoz,
Phys.\ Rev.\ D {\bf 69}, 123502 (2004)
[arXiv:hep-th/0402190];
C.Kim, H.B.Kim, Y.Kim, O.~K.~Kwon and C.O.Lee,
arXiv:hep-th/0404242.

\bibitem{dbi} 
Ashoke Sen,
JHEP {\bf 9910}, 008 (1999) [arXiv:hep-th/9909062];
M. Garousi, Nucl. Phys. B {\bf{584}}, 284 (2000)
               [hep-th/0003122];
              E. Bergshoeff, M. de Roo, T. de Wit, E. Eyras and S. Panda,
JHEP {\bf{0005}}, 009 (2000) [hep-th/0003221];
J. Kluson, Phys. Rev. D {\bf{62}}, 126003 (2000) [hep-th/0004106]

\bibitem{linde} L. Kofman and A. Linde, JHEP, {\bf{07}}, 004 (2002).
                [hep-th/0205121]; L. Kofman, A. Linde and A. Starobinsky

\bibitem{sami} M.R. Garousi, M. Sami and S. Tsujikawa, hep-th/0402075

\bibitem{garousi} M.R. Garousi, JHEP {\bf{0312}}, 035 (2003)

\bibitem{warp} 
K. Dasgupta, G. Rajesh and Sethi, JHEP {\bf 0008}, 023 (1999)
[arXiv: hep-th/9908088]; K. Becker and M. Becker, Nucl. Phys. B
{\bf 477}, 155 (1996) [arXiv: hep-th/9605053]; H. Verlinde, Nucl.
Phys. B {\bf 580}, 264 (2000) [arXiv: hep-th/9906182]; C. Chan,
P. Paul and H. Verlinde, Nucl. Phys. B {\bf 581}, 156 (2000)
[arXiv: hep-th/0003236]; P. Mayr, Nucl. Phys. B {\bf 593}, 99
(2001) [arXiv: hep-th/0003198]; JHEP {\bf 0011}, 013 (2000)
[arXiv: hep-th/0006204]; B. Greene, K. Schalm and G. Shiu, Nucl.
Phys. B {\bf 584}, 480 (2000) [arXiv: hep-th/0004103].
S. Giddings, S. Kachru and J. Polchinski,
Phys. Rev. D {\bf 66}, 106006 (2002) [arXiv: hep-th/0105079].
E. Witten, Nucl. Phys. B {\bf 474}, 343 (1996) [arXiv:
hep-th/9604030].


\bibitem{hwang} J-c Hwang and H. Noh, Phys.Rev.D66:084009, (2002),  
hep-th/0206100

\bibitem{joris} Joris Raeymaekers, hep-th/0406195.
\end{thebibliography}
\end{document}